# Effect of Crystal Quality on HCP-BCC Phase Transition in Solid $^4$He

Nikolay Mikhin, Andrey Polev, Eduard Rudavskii, and Yegor Vekhov

B.Verkin Institute for Low Temperature Physics and Engineering of the National Academy of Sciences of Ukraine, 47 Lenin Ave., Kharkov 61103, UKRAINE
*vekhov@ilt.kharkov.ua*

## Abstract

The kinetics of HCP-BCC structure phase transition is studied by precise pressure measurement technique in $^4$He crystals of different quality. An anomalous pressure behavior in bad quality crystals under constant volume conditions is detected just after HCP-BCC structure phase transition. A sharp pressure drop of 0.2 bar was observed at constant temperature. The subsequent pressure kinetics is a non-monotonic temperature function. The effect observed can be explained if we suppose that microscopic liquid droplets appear on the HCP-BCC interphase region in bad quality crystals. After the interphase region disappearance, these droplets are crystallized with pressure reduction. It is shown that this effect is absent in high quality thermal-treated crystals.

## 1. INTRODUCTION

In recent years the kinetics of phase transition between dramatically different crystalline phases BCC and HCP has been actively studied with the use of various experimental techniques [1-7] - optical, acoustical, NMR, precise barometry, by driving a superconductor wire which is trapped in solid helium. The facts about the possibility of liquid droplet formation or even crystal re-melting under HCP-BCC phase transition are given in numerous works. It is noted that in all cases the kinetics of phase transition is very sensitive to sample prehistory. First of all it means crystal quality and ways of its thermal treatment.

In the present work the method of precise barometry at constant volume is used that allows both to control the crystal quality by pressure magnitude and to study the kinetics of phase transition by pressure variation. Thermo-dynamic and kinetic properties of the crystal during HCP-BCC transition are compared for high quality and not annealed bad quality crystals.

## 2. EXPERIMENTAL TECHNIQUE

The crystal under investigation had a shape of disk 10 mm in diameter and 1.5 mm in thickness. One of the experimental cell walls served as a mobile membrane of the capacitive pressure gauge. The pressure was measured in situ with an accuracy of about 0.003 bar (sensitivity was about 0.001 bar). The sample was connected with the 1K pot via a weak thermal link. The temperature of the crystal was measured with a carbon resistance thermometer with an accuracy of about 5 mK (sensitivity was about 1 mK). The experimental cell is described in more details in Ref. [6].

The crystal sample was grown by the blocking capillary method, so that all the measurements were carried out under constant volume conditions. In the experiments the step-wise heating and cooling of the cold finger were usually realized with a step of 10-15 mK. During this process the kinetics of pressure and temperature in the cell were recorded. The investigations were carried out in a temperature region of 1.3-1.9 K with molar volumes ranging from 20.85 to 21.10 cm$^3$/mol.

3. RESULTS AND DISCUSSIONS

In the work several tens of samples were studied and in all cases the kinetics of HCP-BCC phase transition depended essentially on crystal quality. For the case where the initial state was a mixture of BCC and HCP phases and the crystal was not subjected to any special thermal treatment (annealing or thermo-cycling), for the step-wise heating the typical temperature dependence of pressure P(T) is shown against the background of P-T phase diagram in Fig.1, its kinetics P(t) is shown in Fig.2. On heating the sample by 20 mK (from dot 1 to dot 2) an increase in pressure connected with the total disappearance of a denser HCP phase is observed and then pressure drops by 0.2 bar (from dot 2 to dot 3) at a constant temperature. The following heating of the sample (from dot 3 to dot 4) leads at first to pressure increasing due to thermal expansion of the BCC crystal and then to its decreasing due to sample annealing near the melting curve, as observed in Ref. [7]. The latter supports the fact that the crystal was imperfect.

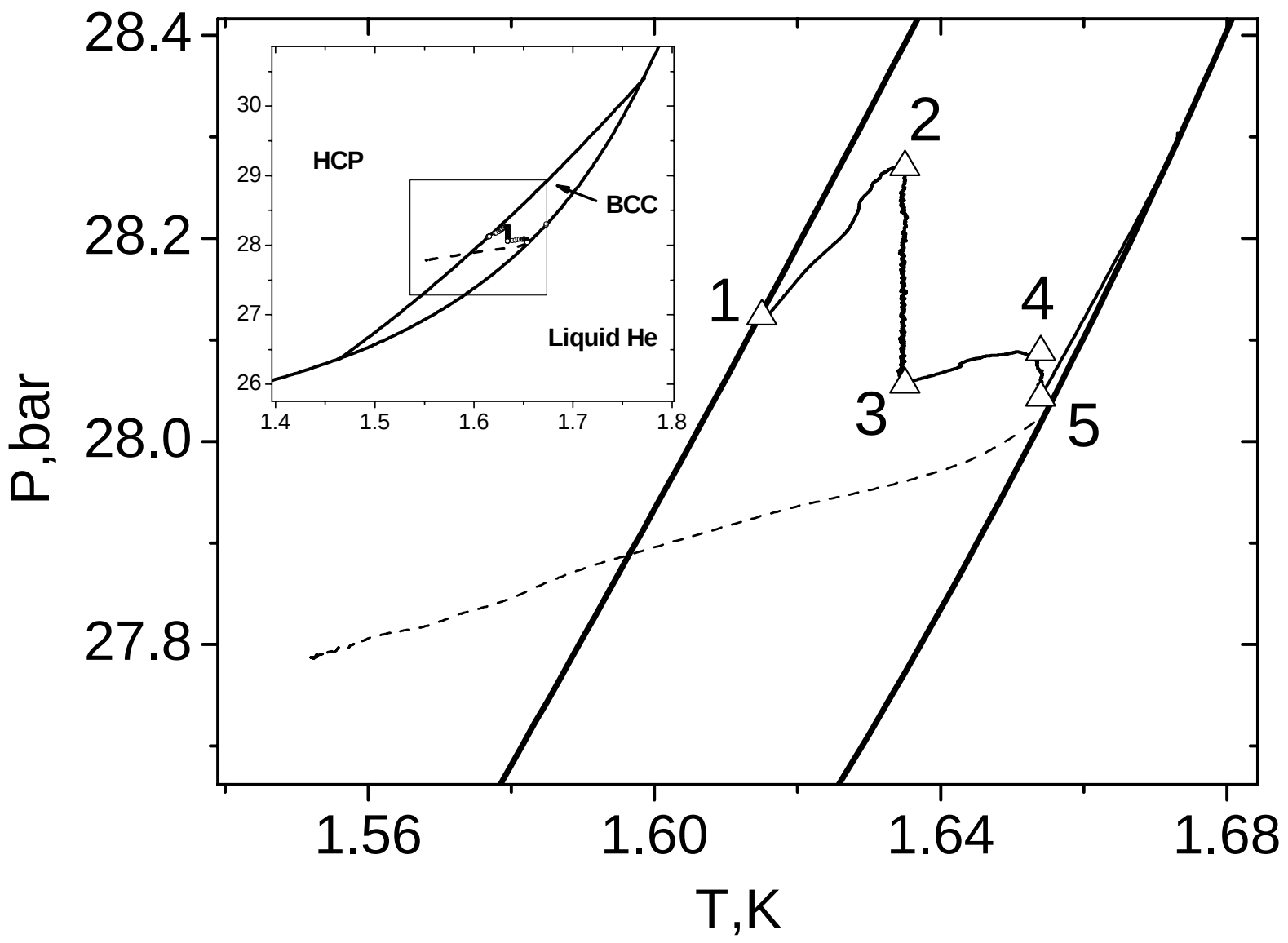

Fig. 1. Pressure variation of sample under step-wise heating and cooling for bad quality crystal (triangles - equilibrium values of P and T under heating; dash line - BCC crystal cooling).

On cooling the BCC crystal, the pressure monotonously decreases with temperature (dash line in Fig.1). In that case one can obtain a much overcooled metastable BCC phase (see Ref. [8]).

It is interesting that the above mentioned unexpected effect of pressure drop after the HCP-BCC transition is observed only in bad quality crystals. To improve the crystal quality a special thermal treatment was used: annealing at the melting curve, annealing in the BCC single phase region near the melting curve and thermo-cycling in single phase regions [7, 8]. The criterion of crystal perfection are, First, continuity of the pressure with time under constant temperature which is closed to melting and, second, reaching of the pressure minimum under thermal cycling.

Similar dependences P(T) and P(t) for HCP-BCC transition in high quality helium crystals are shown in Fig.3 and Fig.4. In this case the dependence P(t) is a monotonous exponential function of time at each heating step. By analogy with the bad quality crystal (Fig.1), a reverse run for cooling the BCC crystal is shown in Fig.3 (dash line). In this case the run coincides with the line of previous heating (with the equilibrium BCC phase), indicating also a good quality of the sample.

The anomalous pressure behavior observed after HCP-BCC transition in helium bad quality crystals can be explained from the following assumption. A poor crystal quality means that sample consists of numerous small crystallites. During the phase transition, while dramatically different crystalline cubic and hexagonal structures are in contact and the interphase boundary is rather large, the regions of local pressure reduction are formed at the BCC-HCP boundary [9]. This can result in a partial crystal melting or liquid droplets formation in the interphase region. As soon as the interphase boundary has disappeared, it is occurred a crystallization of these liquid droplets, resulting in the pressure reduction observed in the experiment (Fig.1, dots 2, 3). The possibility of liquid formation under HCP-BCC transition in solid helium was noted in Ref. [1-3, 5]. The results of these experiments are indirect evidence in the favour of this model and show that it can be realized in helium bad quality crystals. By estimation, pressure decreasing on 0.2 bar (Fig.1) in the crystal with $V_m$=21.0 $cm^3$/mol at 1.63 K can be produced by liquid crystallization of 1% of the sample volume.

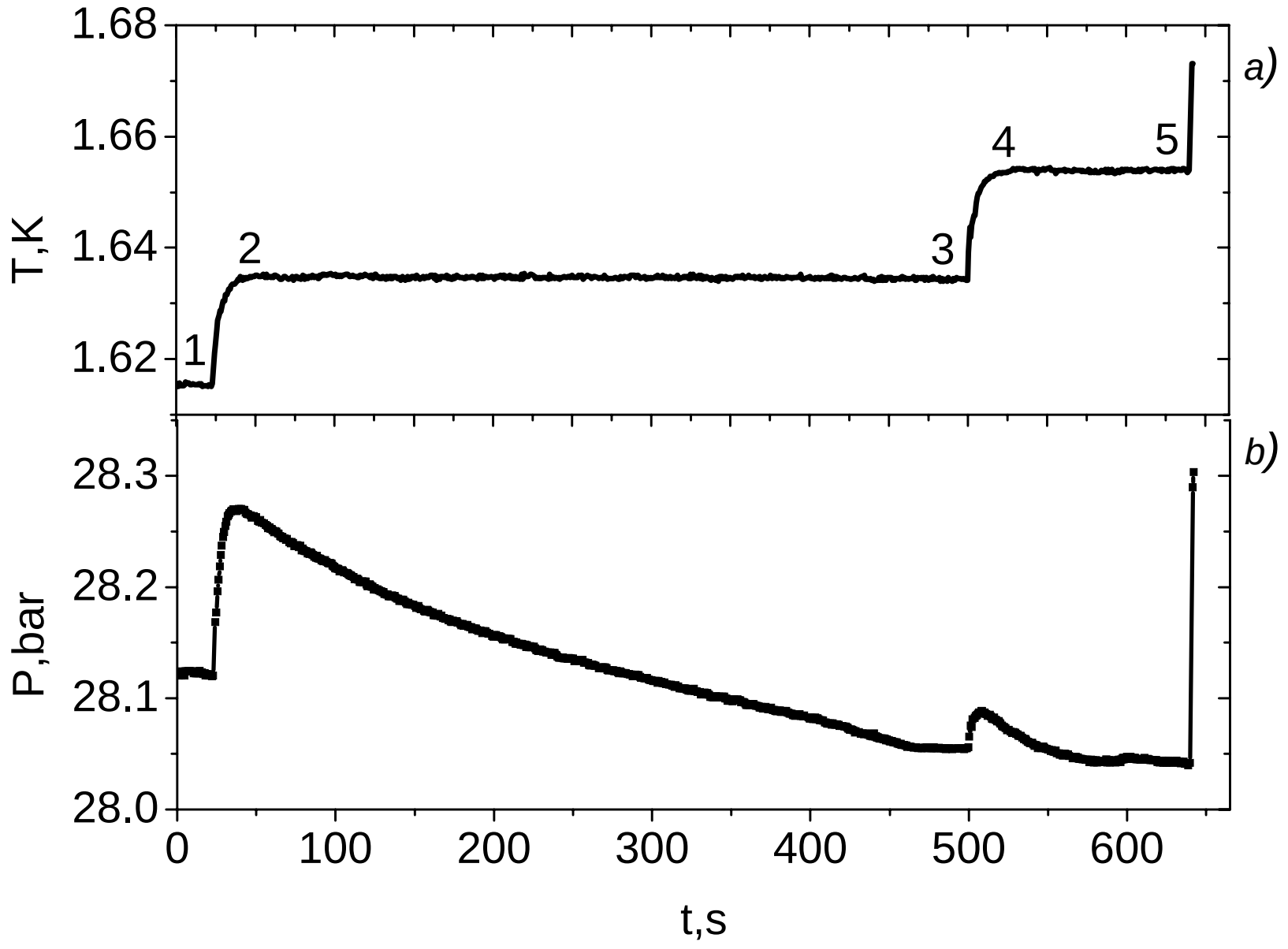

Fig. 2. Kinetics of temperature (a) and pressure (b) variation for bad quality crystal near HCP-BCC phase transition under step-wise heating (numbers correspond to the dots in Fig.1).

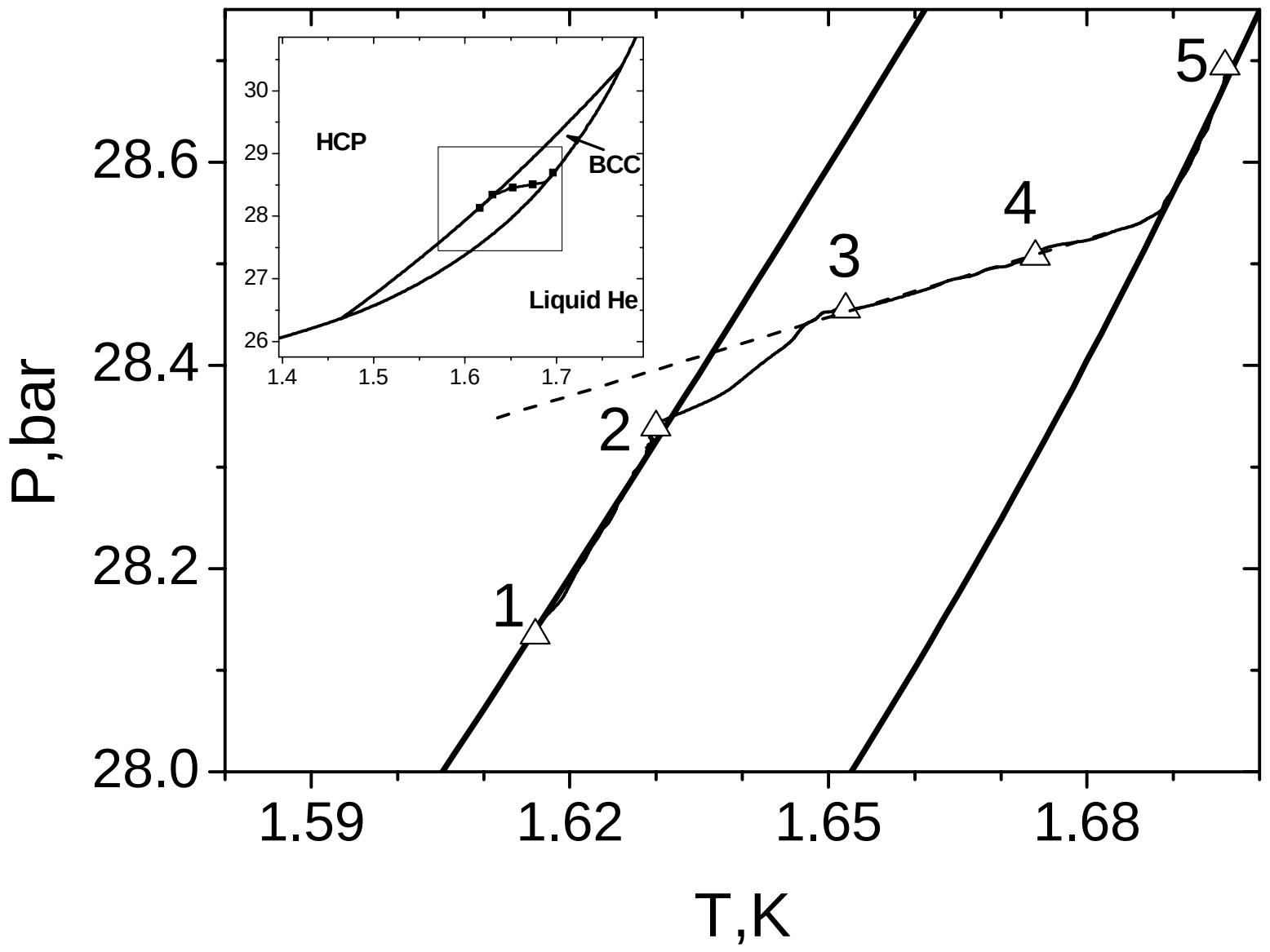

Fig. 3. Pressure variation of sample under step-wise heating and cooling for high quality crystal (triangles - equilibrium values of P and T under heating; dash line - BCC crystal cooling).

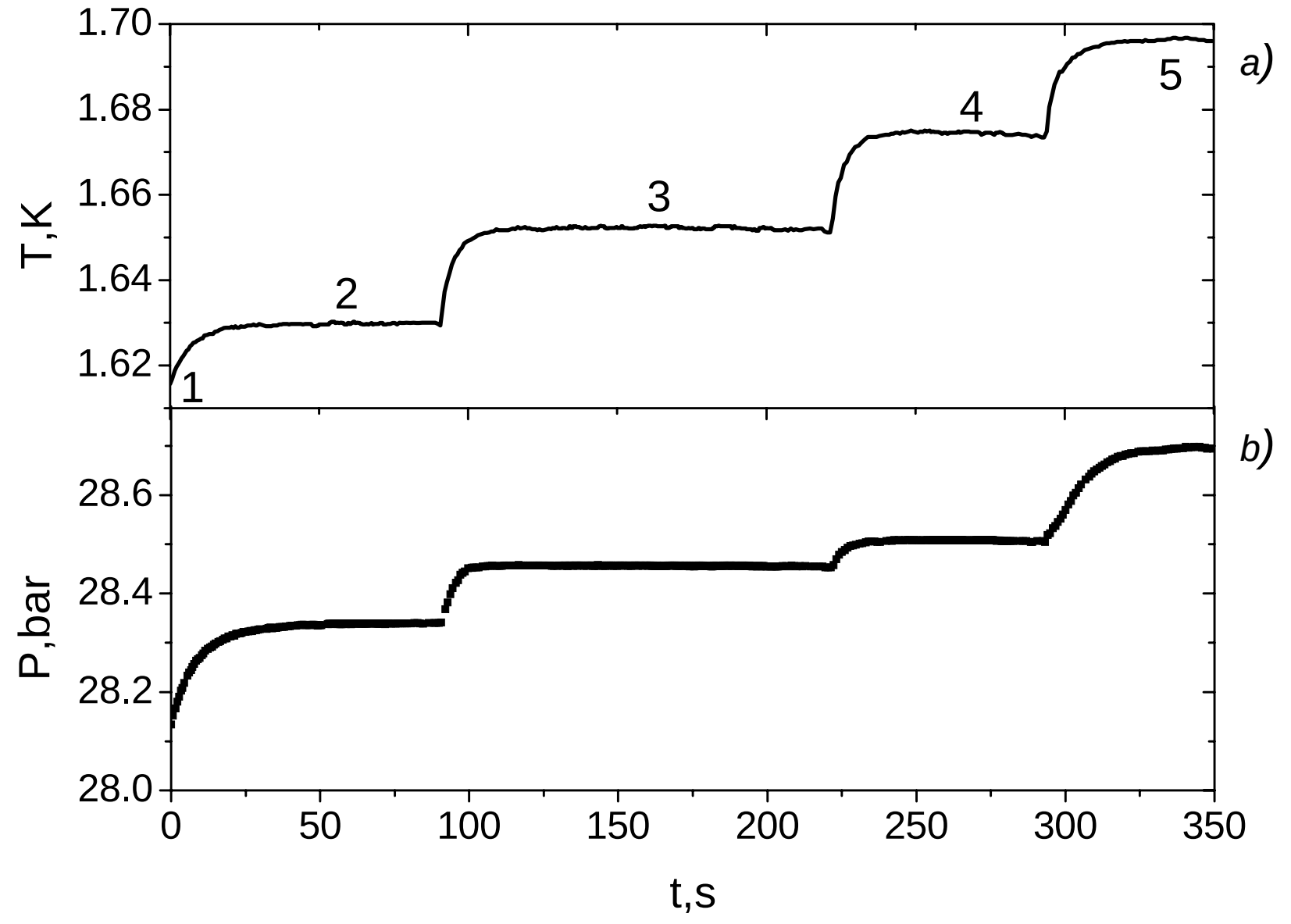

Fig. 4. Kinetics of temperature (*a*) and pressure (*b*) variation for high quality crystal near HCP-BCC phase transition under step-wise heating (numbers correspond to the dots in Fig.3)

The same conclusions follow from our experiments carried out with samples containing 1% of $^3$He.

In the case of reverse BCC-HCP transition, one could expect the same effect. But the experimental data do not confirm this. Possible explanation of this consists in the assumption that the liquid crystallization after the interphase region disappearance cannot lead to pressure reduction below the equilibrium BCC-HCP line. It is realized in continuation of the run along this line.

## ACKNOWLEDGMENTS

We thank A. Birchenco for help in the experiments.

## REFERENCES

1. N.E. Dyumin, V.N. Grigor'ev, and S.V. Svatko, Fiz. Nizk. Temp. (Russ), 15, 253 (1989), Sov. J. Low. Temp. Phys. 15, 142 (1989).
2. I. Berent, and E. Polturak, J. Low Temp. Phys. 112, 337 (1998).
3. E. Polturak, A. Kanigel, N. Gov, T. Markovich, and J. Adler, Physica B 280, 142 (2000).
4. Y. Okuda, H. Fujii, Y. Okumura, and H. Maekawa, J. Low Temp.Phys. 121, 725 (2000).
5. N. Mikhin, A. Polev, and E. Rudavskii, JETP Lett. 73, 470 (2001).
6. Ye. Vekhov, N. Mikhin, A. Polev, and E. Rudavskii, Fiz. Nizk. Temp. (Russ), 31, 1341 (2005), Low Temp. Phys. 31, 1017, (2005).
7. A. Birchenco, Ye. Vekhov, N. Mikhin, A. Polev, and E. Rudavskii, Fiz. Nizk. Temp. (Russ), 32, (2006), Low Temp. Phys, 32, 1471, (2006).
8. N. Mikhin, E. Rudavskii, and Ye. Vekhov, JLTP, to be published.
9. A.I. Karasevskii, Fiz. Nizk. Temp. (Russ), 32, 1203 (2006), Low Temp. Phys., 32, 1203, (2006).